\documentclass{jetpl}
\usepackage[koi8-r]{inputenc}
\usepackage[russian]{babel}
\usepackage{url,color,amsmath,amsfonts,graphics,epsfig,amssymb}
\twocolumn
\lat

\title{%
\textit{Carpet--2} search for gamma rays above 100~TeV in coincidence
with HAWC and IceCube alerts
}

\rtitle{Gamma rays above 100~TeV coincident with HAWC and
IceCube alerts}

\sodtitle{
\textit{Carpet--2} search for gamma rays above 100~TeV in coincidence
with HAWC and IceCube alerts
}

\author{~\\[-7mm]
D.\,D.\,Dzhappuev$^{a}$,
Yu.\,Z.\,Afashokov$^{a}$,
I.\,M.\,Dzaparova$^{a,b}$,
E.\,A.\,Gorbacheva$^{a}$,
I.\,S.\,Karpikov$^{a}$,
M.\,M.\,Khadzhiev$^{a}$,
N.\,F.\,Klimenko$^{a}$,
A.\,U.\,Kudzhaev$^{a}$,
A.\,N.\,Kurenya$^{a}$,
A.\,S.\,Lidvansky$^{a}$,
O.\,I.\,Mikhailova$^{a}$,
V.\,B.\,Petkov$^{a,b}$,
V.\,S.\,Romanenko$^{a}$,
G.\,I.\,Rubtsov$^{a}$,
S.\,V.\,Troitsky$^{a}$\thanks{Corresponding author; e-mail:
st@ms2.inr.ac.ru},
I.\,B.\,Unatlokov$^{a}$,
A.\,F.\,Yanin$^{a}$,
Ya.\,V.\,Zhezher$^{a}$,
K.\,V.\,Zhuravleva$^{a}$\\
~}
\rauthor{Carpet--2 Group}

\sodauthor{D.\,D.\,Dzhappuev et al.\ (Carpet--2 Group)}

\address{
$^{a}$Institute for Nuclear Research of the Russian Academy of
Sciences,\\
60th October Anniversary prospect 7A, 117312 Moscow, Russia\\
$^{b}$Institute of Astronomy, Russian Academy of Sciences, Moscow, 119017
Russia}

\dates{November 4, 2020}{*}

\abstract{We report on the search of astrophysical gamma rays with
energies in the 100~TeV to several PeV range arriving in directional and
temporal coincidence with public alerts from HAWC (TeV gamma rays) and
IceCube (neutrinos above $\sim 100$~TeV). The observations have been
performed with the Carpet--2 air-shower detector at the Baksan Neutrino
Observatory, working in the ``photon-friendly'' mode since 2018. Photon
candidate showers are selected by their low muon content. No
significant excess of the photon candidates have been observed, and upper
limits on gamma-ray fluences associated with the alerts are obtained. For
events with good viewing conditions, the Carpet--2 effective area for
photons is of the order of the IceCube effective area for neutrinos of
the same energy, so the constraints start to probe the production of
neutrinos in fast flares of Galactic sources. }

\begin{document}
\maketitle
\renewcommand{\refname}{}

Over the last decade, multimessenger astronomy has brought several
bright results of great importance, notably the identification of a binary
neutron star merger in gravitational-wave and electromagnetic channels
\cite{GW} and a possible association of a high-energy neutrino event with
a
blazar flare \cite{TXS0506}. These and other important observations became
possible thanks to alerts distributed by gravitational-wave, neutrino
and conventional astronomical observatories to the worldwide community of
observers detecting signals in various channels and electromagnetic bands.
Up to now, in the electromagnetic channel, these alerts have been followed
up at the energies up to $\sim 100$~TeV, above which the sensitivity of
the highest-energy participating observatory, HAWC, fades, while higher
energies have been accessible only for the neutrino and cosmic-ray
channels. The purpose of the present paper is to push further this
high-energy limit for the photon channel and to report the results of the
first ever multimessenger alert follow-up in the 100~TeV to several PeV
gamma-ray band.

This energy band is especially important for high-energy neutrino alerts
because it is precisely the range where the estimated neutrino energies
fall to. The origin of these astrophysical neutrinos remains uncertain,
see e.g.\ Refs.~\cite{rev, VissaniRev} for reviews. Recent studies
indicate \cite{neutradio1, neutradio2} that the entire astrophysical
neutrino flux between 1~TeV and 1~PeV, estimated from muon-track events
and extrapolated to lower energies as a power law \cite{astro-nu-flux},
may
be associated with radio blazars. However, analyses of cascade events
\cite{cascade-flux} indicate that this extrapolation may underestimate the
flux at several tens TeV. The origin of the additional flux component is
unknown, and some Galactic scenarios are discussed \cite{Gal}. Note that
these putative Galactic sources may be distributed isotropically in the
sky because of
constraints on the excess of neutrino emission from the Galactic
plane~\cite{ST-gal, IC-gal}.

In the dominant part of the scenarios
of the astrophysical production of high-energy neutrinos, they are born in
$\pi^{\pm}$-meson decays, while the decays of $\pi^{0}$ mesons result in
the accompanying gamma rays of similar energies. The $\sim (100
\dots 1000)$~TeV photons efficiently produce $e^{\pm}$ pairs on the cosmic
microwave background~\cite{Nikishov}, so the mean free path of these
photons does not exceed the size of the Milky Way. Therefore, photons of
these energies cannot reach us from active galactic nuclei (see
e.g.\ Ref.~\cite{AGN-rev} for a discussion of models), nor from putative
neutrino production regions in the intergalactic space
\cite{Kachelriess-intergalactic, Uryson-intergalactic}. Any observation of
gamma rays at these energies associated with neutrinos would either imply
Galactic sources \cite{Murase-gamma, OK-ST-gamma} or, if the event is
directionally associated with an extragalactic object, suggest new
particle physics affecting the transparency of the Universe to high-energy
gamma rays \cite{ST-axion-rev}.

Previously, we reported \cite{Carpet-IC1} on the analysis of the data
above
1~PeV accumulated by Carpet--2 during 1999--2011, when the main task of
the experiment was related to cosmic-ray studies. We obtained constraints
on the neutrino flux from directions of IceCube neutrinos assuming the
sources are steady, because the data time spans did not overlap. Since
2018, Carpet--2 returned to data taking, now as a gamma-ray observatory:
changes in the trigger and in the data analysis allowed us to significantly
improve the efficiency and to lower the threshold energy for gamma-ray
studies, see e.g.\ Ref.~\cite{Carpet-300TeV}. Here, we report for the
first
time on the results of simultaneous observations of the alerts at
energies above 100~TeV.

IceCube issues public alerts corresponding to detections of individual
muon-track events passing certain criteria since 2017, see the
description of ``EHE'' and ``HESE'' alerts (2017-2019) in Ref.
\cite{EHE-HESE} and of ``GOLD'' and ``BRONZE'' alerts (since 2019) in
Ref.~\cite{GOLD-BRONZE}. Most of the alert events have estimated energies
between 100~TeV and 1~PeV. The criteria are chosen to maximize the
probability that the event is of astrophysical origin. Still, many of them
are background atmospheric events: the astrophysical purity of the
``GOLD'' sample is only 50\%, that of the ``BRONZE'' sample is 30\%.

Another window into sub-PeV astrophysics is provided by observations of
gamma rays at slightly lower energies in the TeV range. In parallel with
pointing observations by atmospheric Cerenkov telescopes, air-shower
installations monitor the sky continuously. In particular, HAWC
has recently started
to issue various public alerts when a significant point-like signal is
observed during one dayly passage of the ``flare'' direction through the
HAWC field of view, see Ref.~\cite{HAWCalerts} for more details. These
alerts are of particular interest because the energies are very close to
the band we study here, and a discovery of a flaring Galactic source in
this energy range would have important astrophysical consequences.

We report here on the results of 2.5 years of following up these HAWC
and IceCube alerts with Carpet--2, see Refs.~\cite{0902.0252, 1511.09397}
for the description of the experiment. Since Carpet--2 is an air-shower
detector, it operates continuously and observes a large fraction of the
sky. Therefore it does not need to be pointed to the alert direction, so
the alerts are analysed offline. The search is based on the set of photon
candidate events which are selected as muon-poor air showers, as described
in Refs.~\cite{Carpet-IC1, Carpet-300TeV, Dzhappuev:2018bnl}. The photon
detection efficiency and the angular resolution were determined by the
Monte-Carlo simulations; together with the photon-candidate selection cuts
for the energies $E>300$~TeV they are discussed in detail in
Ref.~\cite{Carpet-300TeV} (``Dataset~II''). Here, we extend the analysis
down to 100~TeV energies because the combined directional and temporal
selection reduces efficiently the hadronic background, which otherwise is
the main problem in the search for gamma rays with Carpet--2 below
300~TeV. Photon candidate events are selected by their reconstructed
number of charged particles, $N_{e}$, and the number of muons in the
175~m$^{2}$ muon detector, $n_{\mu}$. It was found optimal to consider
only muonless events for these energies, though at higher energies the cut
was determined in terms of the ratio $n_{\mu}/N_{e}$. The criteria are
given in Table~\ref{tab:crit}
\begin{table}
\begin{center}
\begin{tabular}{ccccc}
\hline
\hline
Energy,   & Min.  & Max.  & Selection  & Number
\\
TeV       & $N_{e}$ & $n_{\mu}/N_{e}$ & efficiency & of events\\
\hline
$>$100 &    $10^{4.2251}$  &   0   &   0.995   &1021\\
$>$300 &    $10^{4.6372}$  &   0   &   0.509   & 598\\
\hline
\hline
\end{tabular}
\end{center}
\caption{\label{tab:crit}
Table~\ref{tab:crit}:
Two samples of photon candidate events.
}
\end{table}
together with
the number of photon candidate events in the data sample and the efficiency
of the gamma-ray selection cuts (determined as the ratio of Monte-Carlo
photons with reconstructed $n_{\mu}$, $N_{e}$ satisfying the selection
conditions to the total number of reconstructed Monte-Carlo photons). The
candidates were selected from 52791 air showers, recorded between April 8,
2018, and October 26, 2020, succesfully reconstructed and passing the
quality cuts. The number of live days in this period was 675.

Carpet--2 dataset consists of events being observed with zenith angles up
to 40$^{\circ}$, but the efficiency decreases fastly for inclined showers.
We select the alerts with declinations between $+5^{\circ}$ and
$+76^{\circ}$ so that the maximal elevation corresponds to a zenith angle
not exceeding $35^{\circ}$. In addition, we drop the events arrived at
the days when our data were not recorded because of maintenance. In this
way we arrive to the list of 9 HAWC alerts and 22 IceCube alerts presented
in Table~\ref{tab:IC} in the Supplementary Material~\cite{SI}.

Most of the alert directions, however, were outside the Carpet--2 field of
view at the moment of the event, but they passed through the field of view
during the day. We therefore determine time windows of 24~hours and of
30~days, centered at the event moment, to search for coinciding gamma-ray
candidates. The angular window, as discussed before~\cite{Carpet-300TeV},
was set to the 90\% CL angular resolution, which at these energies is
about 6.15$^{\circ}$. The expected number of candidate events was
calculated by randomizing arrival times of photon candidates in the
sample.

For each particular alert we present the expected and observed
numbers of the events, the estimated flux within the selected time
window and the fluence in Tables~\ref{tab:100}, \ref{tab:300} in the
Supplementary Material~\cite{SI}. Note that the flux estimates depend
crucially on the time window, which is chosen more or less arbitrarily,
while the most interesting physical quantity for a burst is the fluence. No
significant excess of photon candidates was found, and we present 95\% CL
upper limits on the flux and fluence.  These limits vary strongly from
one alert to another because of the strong dependence of the
reconstruction efficiency for photons on the zenith angle. In all cases we
are close to the ``zero signal, zero background'' regime which motivates
us to use stacking to improve sensitivity, and stacked results for all
HAWC alerts and all IceCube alerts are also presented in the same tables.

A weak, two-sigma excess, 8 events observed for 4.25 expected, is found
at $E>100$~TeV for stacked IceCube events, dominated by the 200911A
alert. This excess is consistent with expected fluctuations, given
multiple trials.

Directions of two IceCube events, 190331A and 191215A, were in the
Carpet--2 field of view at the moment of neutrino arrivals, which allows
us to estimate directly their fluence assuming a fast flare. We use the
time window of 1000~s for these two alerts. For good viewing conditions,
that is small zenith angles, the effective area of Carpet--2 in the
configuration used for the photon search in the present analysis is of the
same order as the effective area of IceCube for neutrinos of similar
energies, see Fig.~\ref{fig:effareas}.
\begin{figure}
\centering
\includegraphics[width=0.96\columnwidth,clip]{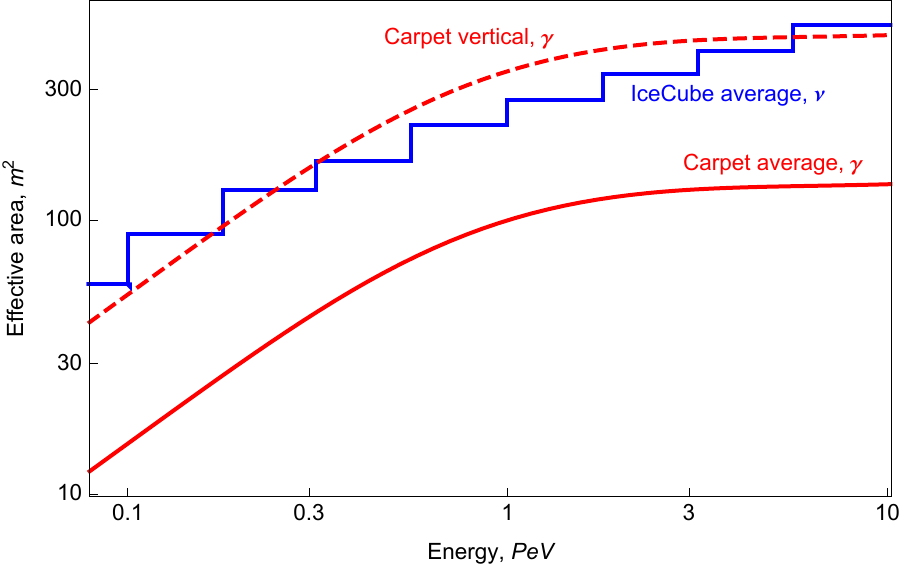}
\caption{\label{fig:effareas}
Figure~\ref{fig:effareas}: Comparison of effective areas of Carpet--2 (red
continuous lines, photon detection, present analysis) and
IceCube (blue step line, neutrino detection, muon tracks, average over the
Northern hemisphere \cite{ICeffA}). For Carpet--2, the full line gives the
average over the field of view while the dashed line corresponds to
vertical events. }
\end{figure}
Therefore, the fluence upper
limits of order $\lesssim 1$~GeV/cm$^{2}$, see
Tables~\ref{tab:100}, \ref{tab:300}, \ref{tab:1000sec} in the Supplemental
material~\cite{SI}, start to constrain the origin of neutrinos in fast
flares of Galactic sources. Assuming flares longer than a few hours, we
obtain less constraining limits because the arrival direction of the alert
event leaves the field of view of Carpet--2. All HAWC alerts happen
outside of the Carpet--2 field of view because of the difference in
geographical longuitudes of the two installations.

The program of multimessenger observations with Carpet--2 continues.
Besides HAWC and IceCube alerts, it includes also LIGO/VIRGO alerts and
low-energy neutrino burst alerts from the Baksan Underground Scintillator
Telescope. Soon, when Baikal-GVD high-energy neutrino alerts become
available, they will join the list. At the same time, the upgrade of the
installation to Carpet--3 is ongoing with an order of magnitude increase in
the collection area and more than a twofold increase in the area of the
muon detector. This will allow to reach the sensitivity in the sub-PeV
gamma rays at the level of the corresponding neutrino sensitivity of
IceCube not only for short flares, but also for long-term observations and
for the diffuse flux.

This work is supported in the framework of the State project ``Science''
by the Ministry of Science and Higher Education of the Russian Federation
under the contract 075-15-2020-778.

%
\lat
%

\title{%
Supplemental material to\\
``\textit{Carpet--2} search for gamma rays above 100~TeV in coincidence
with HAWC and IceCube alerts''}

\author{~\\[-7mm]
D.\,D.\,Dzhappuev et al.\ (Carpet--2 Group)
~}
\abstract{}

\maketitle

This Supplemental Material presents results of the search for $E>100$~TeV
photons associated with individual alerts. Table~\ref{tab:IC} gives the
list of alerts, Tables \ref{tab:100} and \ref{tab:300} present the numbers
of expected and observed (in 24-hour and 30-day time windows)
photon-candidate events and upper limits on fluxes and fluences for
$E>100$~TeV and $E>300$~TeV, respectively. Table \ref{tab:1000sec}
presents the same quantities for 1000-second time windows for two alerts
whose arrival direction was in the Carpet--2 field of view at the event
arrival time.

\begin{table*}
\begin{center}
\begin{tabular}{lllrr}
\hline
\hline
type   &  ID   & MJD    & RA, $^{\circ}$ & DEC, $^{\circ}$ \\
\hline
HAWC &  H190806A & 58701.556 &    78.4 &  6.6 \\
HAWC &  H190917A & 58743.052 &   321.8 & 40.0 \\
HAWC &  H190927A & 58753.954 &   248.7 & 21.2 \\
HAWC &  H191019A & 58775.841 &   217.5 & 25.8 \\
HAWC &  H191208A & 58825.557 &   228.9 & 40.2 \\
HAWC &  H200314A & 58922.446 &   255.7 & 48.1 \\
HAWC &  H200814A & 59075.901 &   177.8 & 19.9 \\
HAWC &  H200815A & 59076.264 &    11.2 & 11.5 \\
HAWC &  H201019A & 59141.905 &   203.1 & 29.7 \\
HESE &  I190331A & 58573.289 &   355.6 & 71.1 \\
EHE  &  I190503A & 58606.724 &   120.3 &  6.4 \\
GOLD &  I190619A & 58653.552 &   343.3 & 10.7 \\
GOLD &  I190730A & 58694.869 &   225.8 & 10.5 \\
GOLD &  I191001A & 58757.840 &   314.1 & 12.9 \\
GOLD &  I200109A & 58857.987 &   164.5 & 11.9 \\
GOLD &  I200530A & 58999.330 &   254.4 & 27.5 \\
BRONZE& I190704A & 58668.784 &   161.8 & 27.1 \\
BRONZE& I190712A & 58676.052 &    76.5 & 13.1 \\
BRONZE& I191215A & 58832.465 &   285.9 & 58.9 \\
BRONZE& I191231A & 58848.458 &    46.4 & 20.4 \\
BRONZE& I200117A & 58865.464 &   116.2 & 29.1 \\
BRONZE& I200410A & 58949.972 &   241.3 & 11.6 \\
BRONZE& I200425A & 58964.977 &   100.1 & 53.6 \\
BRONZE& I200512A & 58981.314 &   295.2 & 15.8 \\
BRONZE& I200614A & 59014.529 &    33.8 & 31.6 \\
BRONZE& I200620A & 59020.127 &   162.1 & 11.9 \\
BRONZE& I200911A & 59103.597 &    51.1 & 38.1 \\
BRONZE& I200916A & 59108.861 &   109.8 & 14.4 \\
BRONZE& I200921A & 59113.797 &   195.3 & 26.2 \\
BRONZE& I201014A & 59136.093 &   221.2 & 14.4 \\
BRONZE& I201021A & 59143.276 &   260.8 & 14.6 \\
\hline
\hline
\end{tabular}
\end{center}
\caption{\label{tab:IC}
Table~\ref{tab:IC}:
List of alerts used in the present analysis.
HAWC alerts are published at
\protect\url{https://gcn.gsfc.nasa.gov/amon_hawc_events.html}, HESE
IceCube alerts were published at
\protect\url{https://gcn.gsfc.nasa.gov/amon_hese_events.html}, EHE IceCube
alerts were published at
\protect\url{https://gcn.gsfc.nasa.gov/amon_ehe_events.html}, GOLD and
BRONZE IceCube alerts are published at
\protect\url{https://gcn.gsfc.nasa.gov/amon_icecube_gold_bronze_events.html}.
We add ``H'' for HAWC or ``I'' for IceCube to the official alert ID for
clarity. MJD gives the event time as a Modified Julian Day,
R.A.\ and DEC are equatorial coordinates in degrees.
}
\end{table*}

\begin{table*}
\begin{center}
\textbf{$\mathbf{E>100}$~TeV limits}\\[6pt]
\begin{tabular}{ccccccccc}
\hline
\hline
        &  \multicolumn{3}{c}{24 hours} &~~& \multicolumn{4}{c}{30 days}
\\
\cline{2-4}
\cline{6-9}
ID & $n_{\rm expected}$ & $F_{95}$, &
                                 $\mathcal{F}_{95}$,  & &
           $n_{\rm observed}$
          & $n_{\rm expected}$ & $F_{95}$, &
                                 $\mathcal{F}_{95}$, \\
& & 10$^{-11}$cm$^{-2}$s$^{-1}$ & GeV/cm$^{2}$ & & & & 10$^{-11}$cm$^{-2}$s$^{-1}$ &
GeV/cm$^{2}$ \\
\hline
H190806A   &0.0009 &1790.  &154.  &  & 0     &0.0262 &59.    & 153. \\
H190917A   &0.0170 &49.9   &4.31  &  & 1     &0.5093 &2.37   & 6.13 \\
H190927A   &0.0060 &172.   &14.9  &  & 2     &0.1809 &11.7   & 30.4 \\
H191019A   &0.0092 &109.   &9.44  &  & 0     &0.2747 &3.32   & 8.6  \\
H191208A   &0.0167 &49.7   &4.3   &  & 0     &0.5013 &1.39   & 3.6  \\
H200314A   &0.0179 &47.9   &4.14  &  & 0     &0.5360 &1.32   & 3.42 \\
H200814A   &0.0052 &199.   &17.2  &  & 0     &0.1556 &6.31   & 16.4 \\
H200815A   &0.0021 &650.   &56.2  &  & 0     &0.0636 &21.2   & 55.  \\
H201019A   &0.0113 &80.4   &6.95  &  & 1     &0.3396 &3.96   & 10.3 \\
\hline
HAWC stacked&0.0862&  10.2  &   0.88  & &4 &   2.5871 &   0.77  &1.99\\
\hline
\hline
I190331A   &0.0054 &201.   &17.4  &  & 0     &0.1618 &6.35   & 16.5 \\
I190503A   &0.0007 &1890.  &163.  &  & 0     &0.0222 &62.4   & 162. \\
I190619A   &0.0019 &747.   &64.6  &  & 1     &0.0582 &39.    & 101. \\
I190730A   &0.0017 &774.   &66.9  &  & 0     &0.0516 &25.4   & 65.8 \\
I191001A   &0.0024 &517.   &44.7  &  & 0     &0.0720 &16.8   & 43.7 \\
I200109A   &0.0021 &609.   &52.6  &  & 1     &0.0618 &31.7   & 82.3 \\
I200530A   &0.0106 &94.8   &8.19  &  & 0     &0.3182 &2.83   & 7.34 \\
I190704A   &0.0104 &97.9   &8.46  &  & 0     &0.3111 &2.93   & 7.61 \\
I190712A   &0.0024 &501.   &43.3  &  & 0     &0.0733 &16.3   & 42.3 \\
I191215A   &0.0134 &70.8   &6.12  &  & 0     &0.4013 &2.05   & 5.32 \\
I191231A   &0.0057 &188.   &16.2  &  & 1     &0.1702 &9.59   & 24.8 \\
I200117A   &0.0109 &83.9   &7.25  &  & 2     &0.3262 &5.59   & 14.5 \\
I200410A   &0.0019 &639.   &55.2  &  & 0     &0.0556 &20.9   & 54.2 \\
I200425A   &0.0167 &54.9   &4.74  &  & 0     &0.5018 &1.53   & 3.97 \\
I200512A   &0.0037 &336.   &29.   &  & 0     &0.1116 &10.8   & 28.  \\
I200614A   &0.0136 &71.    &6.14  &  & 0     &0.4080 &2.05   & 5.32 \\
I200620A   &0.0022 &609.   &52.6  &  & 0     &0.0667 &19.9   & 51.5 \\
I200911A   &0.0168 &52.6   &4.54  &  & 3     &0.5027 &4.27   & 11.1 \\
I200916A   &0.0029 &410.   &35.5  &  & 0     &0.0862 &13.3   & 34.5 \\
I200921A   &0.0100 &106.   &9.12  &  & 0     &0.2996 &3.18   & 8.23 \\
I201014A   &0.0029 &410.   &35.5  &  & 0     &0.0880 &13.3   & 34.5 \\
I201021A   &0.0034 &399.   &34.4  &  & 0     &0.1022 &12.8   & 33.3 \\
\hline
IceCube stacked&0.1417&  6.82  &  0.59  & &8  &   4.2502 &   0.81  &2.10\\
\hline
\hline
\end{tabular}
\end{center}
\caption{\label{tab:100}
Table~\ref{tab:100}:
Results of the search of $E>100$~TeV photons. Alert IDs are defined in
Table~\ref{tab:IC}. For a given time period, 24~hours or 30~days centered
at the alert, $n_{\rm expected}$ and $n_{\rm observed}$ are the expected
and observed numbers of photon-like events from the alert direction,
respectively ($n_{\rm observed} =0$ for all cases for 24~hours); $F_{95}$
and $\mathcal{F}_{95}$ are the 95\% CL upper limits on the flux and
fluence of the putative burst associate with the alert.}
\end{table*}

\begin{table*}
\begin{center}
\textbf{$\mathbf{E>300}$~TeV limits}\\[6pt]
\begin{tabular}{ccccccccc}
\hline
\hline
        &  \multicolumn{3}{c}{24 hours} &~~& \multicolumn{4}{c}{30 days}
\\
\cline{2-4}
\cline{6-9}
ID & $n_{\rm expected}$ & $F_{95}$, &
                                 $\mathcal{F}_{95}$,  & &
           $n_{\rm observed}$
          & $n_{\rm expected}$ & $F_{95}$, &
                                 $\mathcal{F}_{95}$, \\
& & 10$^{-11}$cm$^{-2}$s$^{-1}$ & GeV/cm$^{2}$ & & & & 10$^{-11}$cm$^{-2}$s$^{-1}$ &
GeV/cm$^{2}$ \\
\hline
H190806A &   0.0006&  1350. &   351.  & &0  &   0.0182 &   44.9  &349.\\
H190917A &   0.0102&  38.   &   9.84  & &1  &   0.3053 &   1.88  &14.6\\
H190927A &   0.0033&  131.  &   33.9  & &2  &   0.0982 &   9.02  &70.1\\
H191019A &   0.0049&  83.   &   21.5  & &0  &   0.1480 &   2.64  &20.5\\
H191208A &   0.0102&  37.8  &   9.8   & &0  &   0.3058 &   1.14  &8.83\\
H200314A &   0.0119&  36.4  &   9.44  & &0  &   0.3564 &   1.07  &8.35\\
H200814A &   0.0031&  151.  &   39.2  & &0  &   0.0929 &   4.89  &38. \\
H200815A &   0.0014&  493.  &   128.  & &0  &   0.0427 &   16.2  &126.\\
H201019A &   0.0065&  61.1  &   15.8  & &0  &   0.1964 &   1.91  &14.8\\
\hline
HAWC stacked&0.0521&  7.84  &   2.03  & &3  &   1.5640 &   0.55  &4.27\\
\hline
\hline
I190331A &   0.0030&  153.  &   39.5  & &0  &   0.0889 &   4.94  &38.4\\
I190503A &   0.0005&  1430. &   371.  & &0  &   0.0160 &   47.5  &369.\\
I190619A &   0.0014&  567.  &   147.  & &1  &   0.0413 &   29.7  &231.\\
I190730A &   0.0010&  588.  &   152.  & &0  &   0.0311 &   19.4  &151.\\
I191001A &   0.0015&  393.  &   102.  & &0  &   0.0440 &   12.9  &100.\\
I200109A &   0.0014&  462.  &   120.  & &0  &   0.0427 &   15.2  &118.\\
I200530A &   0.0055&  72.   &   18.7  & &0  &   0.1640 &   2.27  &17.7\\
I190704A &   0.0054&  74.4  &   19.3  & &0  &   0.1631 &   2.35  &18.3\\
I190712A &   0.0013&  380.  &   98.6  & &0  &   0.0387 &   12.5  &97.4\\
I191215A &   0.0083&  53.8  &   14.   & &0  &   0.2480 &   1.65  &12.8\\
I191231A &   0.0034&  143.  &   37.   & &1  &   0.1009 &   7.38  &57.4\\
I200117A &   0.0059&  63.8  &   16.5  & &0  &   0.1778 &   2.    &15.6\\
I200410A &   0.0012&  485.  &   126.  & &0  &   0.0347 &   16.   &124.\\
I200425A &   0.0104&  41.7  &   10.8  & &0  &   0.3111 &   1.25  &9.73\\
I200512A &   0.0019&  255.  &   66.2  & &0  &   0.0582 &   8.35  &64.9\\
I200614A &   0.0070&  54.   &   14.   & &0  &   0.2111 &   1.68  &13. \\
I200620A &   0.0012&  462.  &   120.  & &0  &   0.0364 &   15.2  &118.\\
I200911A &   0.0095&  40.   &   10.4  & &1  &   0.2836 &   1.99  &15.5\\
I200916A &   0.0019&  312.  &   80.8  & &0  &   0.0564 &   10.2  &79.3\\
I200921A &   0.0054&  80.2  &   20.8  & &0  &   0.1609 &   2.53  &19.7\\
I201014A &   0.0016&  312.  &   80.8  & &0  &   0.0476 &   10.2  &79.5\\
I201021A &   0.0020&  303.  &   78.4  & &0  &   0.0587 &   9.89  &76.9\\
\hline
IceCube stacked&0.0805&  5.28  &   1.37  & &3  &   2.4151 &   0.32  &2.51\\
\hline
\hline
\end{tabular}
\end{center}
\caption{\label{tab:300}
Table~\ref{tab:300}:
Results of the search of $E>300$~TeV photons.
See Table~\ref{tab:100} for notations.
}
\end{table*}

\begin{table*}
\begin{center}
\textbf{1000-second limits}\\[6pt]
\begin{tabular}{cccccc}
\hline
\hline
        &  \multicolumn{2}{c}{$E>100$~TeV} &~~&
\multicolumn{2}{c}{$E>300$~TeV} \\
\cline{2-3}
\cline{5-6}
ID & $F_{95}$, & $\mathcal{F}_{95}$,  & & $F_{95}$, &$\mathcal{F}_{95}$, \\
&  10$^{-11}$cm$^{-2}$s$^{-1}$ & GeV/cm$^{2}$ & &
10$^{-11}$cm$^{-2}$s$^{-1}$ & GeV/cm$^{2}$ \\
\hline
I190331A &    7124.  &   7.12  & &   5405.  &16.2\\
I191215A &    1065.  &   1.07   & &   808.   &2.43\\
\hline
\hline
\end{tabular}
\end{center}
\caption{\label{tab:1000sec}
Table~\ref{tab:1000sec}:
Results for alerts in the field of view (1000~s time window). See
Table~\ref{tab:100} for notations.
}
\end{table*}


\end{document}